\documentclass[12pt,preprint]{aastex}

\usepackage{lscape}
\usepackage{epstopdf}

\shorttitle{Pericenter Outburst}
\shortauthors{R.E. Wilson and E.J. Devinney}
\begin{document}

\title{Lobe Overflow as the Likely Cause of Pericenter Outburst in an SMBH Orbiter}

\author{R.E. Wilson\altaffilmark{1,2} \& E. J. Devinney, Jr.\altaffilmark{3}}

\altaffiltext{1}{Astronomy Department, University of Florida, Gainesville, FL 32611 E-mail: rewilson@ufl.edu}

\altaffiltext{2}{Astronomy Department, Indiana University, Swain Hall West, Bloomington, IN 47405}

\altaffiltext{3}{Department of Astrophysics and Planetary Science, Villanova University, 800 East Lancaster Avenue, Villanova, Pa., 19085, USA E-mail: edward.devinney@villanova.edu}

\begin{abstract}

A very large lobe overflow event is suggested to explain the $0.^m4$ brightening observed in K band at pericenter passage of the
star known as S2 that orbits the Galaxy's supermassive black hole (SMBH).
Known observed properties of S2 that contribute to lobe filling are 1) the enormous mass ratio,
$M_{SMBH}/M_{S2}$, 2) S2's fast rotation, and 3) S2's large orbital eccentricity.
Published estimates have given limiting lobe sizes of order 100 to 300 $R_\odot$ but, 
with S2's fast rotation taken into account, the computed lobe size is \textit{much} 
smaller, being compatible with either a main sequence OB star or a stripped evolved star.
An important evolutionary consideration that predicts very large pericenter overflows is envelope 
expansion following mass loss that is characteristic of highly evolved stars.
Material removed by lobe overflow at pericenter is replenished by envelope expansion
as an evolved star awaits its next pericenter passage.
An observational signature of lobe overflow for upcoming pericenter passages would be appearance of
emission lines as the ejected gas expands and becomes optically thin.

\end{abstract}


\section{Background}

Prospects for understanding the evolution of the Galactic center region were
boosted by discovery and quantitative investigation \citep{schodel}
of a star designated as S2 that is bound to the Galaxy's
supermassive black hole (SMBH) in a 15.8 year orbit, with additional
stars in the region under similar investigation -- see also \citet{ghez}.
Many papers have since added observations and ideas on individual SMBH orbiters
and on the statistics and collective properties of stars in the inner
Galactic center region,  e.g. \citet{ghezbeck,ghezduch,alex,blum,eisenhauer,gill09,davies,gill13,witzel}
and an extensive review by \citet{genzel}.

Issues abound regarding the evolutionary states of these stars and how they came to be in the
inner Galactic center, e.g. \citet{davies,zhang,gill09,genzel}. 
Stars arriving on nearly parabolic or even hyperbolic orbits could suffer major 
stripping on initial pericenter passage \citep{davies}, with the lost material carrying away enough orbital energy to leave the
remnant in an elliptical orbit, filling its limiting lobe at pericenter.
Alternatively, a star that has been trapped into a tight orbit while
well detached from its lobe could later undergo evolutionary expansion and attain lobe filling.
Perhaps a lost binary companion may carry off the requisite orbital energy at first encounter with the SMBH and leave
the remaining star bound.

\citet{eisenhauer} found most of the brighter inner orbiters to be B0 to B9 main sequence  
stars, with S2 in the range O8 to B0, and with rotation velocities typical of B stars in the Galactic disk.
\citet{davies} argued that they are actually tidally stripped remnants of 
AGB stars that now superficially resemble main sequence stars and estimated S2's mass at below 
a solar mass, specifically about 0.8 $M_\odot$. All in all, mass estimates for S2 that have been
published or correspond to observed (main sequence) spectral types range from 0.8
to more than 20 $M_\odot$. Accordingly, we simply adopt 10 $M_\odot$ for exploratory
lobe size computations.

\section{Brightening of Star S2 Due to Lobe Overflow} \label{s2_sec}

S2 has continued as the most thoroughly discussed SMBH orbiter, largely due
to its having been observed over more than a full orbit, and 
having brightened by about $0.^m40$ in K band, coincident with pericenter passage \citep{gill09}.
\citeauthor{gill09} offered seven ideas for explaining the brightening,
however they then ruled out four of the ideas and argued against likelihood of the other three.
Consequences of lobe overflow -- a very common player in a variety of close binary issues --
were not among the seven ideas. 
Lobe overflow might not be considered if one were thinking only of the huge orbital scale (of order 1000 AU)
compared to the size of a main sequence star, but it turns out that S2's limiting lobe is 
actually similar in size to a main sequence star of, say, 10 $M_\odot$, as shown in \S\ref{loberad}.
Lobe overflow is an attractive idea for S2 brightening -- one could postulate that S2 exceeded its
limiting lobe at the 2002 pericenter passage and ejected a strong puff of material that quickly
expanded in vacuum so as to appear as a rather large cloud of brightly emitting gas.
The lobe size issue will now be addressed.

\section{Lobe Size Essentials} \label{lobesize}

To place the present work in context, we review the formal relations in estimates for
(1) tidal radii and (2) limiting lobes. Although these two terms are quite 
distinct, both often go by the name 'Roche limit' and some
recent papers treat them as equivalent, thereby leading to considerable confusion and perhaps even
wrong conclusions.  
Item 1, tidal radius, concerns the distance from a mass at which an idealized fluid mass (usually a small
satellite) is disrupted when tidal stretching matches or exceeds the satellite's cohesion due to self-gravity.
The simple tidal radius concept considers
test particles on the surface of a self-gravitating sphere of mass $m$ and radius $r$ at a
distance $d$ from an object of mass $M$. The test particles, located on opposite ends of a diameter of $m$ on the line of
centers, suffer a stretching force (surface to center) per unit particle mass due to external mass $M$ of $2GMr/d^3$, assuming $r$ to be very small.
The effective compressional force per unit particle mass between surface and center that can result in a static
configuration is the object's surface gravity, $Gm/r^2$.
Quantity $d$ is the tidal radius and marks the distance from $M$ at which a very small object
is disrupted, so the final relation pertains to the test particles being arbitrarily close together.
If $r$ is not small compared to $d$, then the simple relation may still give the tidal radius
approximately, with only gravitation considered, although the full relation is then more complicated. A flaw in this picture with regard to
stretched objects of finite size is that the satellite is presumed spherical, whereas tidally stretched
stars lack front-to-back symmetry (i.e. have "teardrop" shapes).
But more important for
the case of star S2 and probably other SMBH orbiters is that rotation of $m$ is not considered in the traditional tidal limit development.
In summation, the tidal limit relation between $d$ and $r$, $d=r(2M/m)^{1/3}$, can be inverted with $d$ set to the orbital separation to give
roughly correct limiting size where the assumptions apply, namely \textit{for small, synchronously rotating\footnote{In this context, 'synchronous rotation'
means that angular star rotation and mean orbital angular rotation are equal or, equivalently, that the star rotates once per orbit period. Note that there are
other meanings of 'synchronous rotation', each valid in its own context.} satellites},
but may be wrong by orders of magnitude for fast rotating stars such as S2.

Item 2 is commonly called a Roche lobe, although it was not originated or even considered by 
Roche, who did however consider a special case of the potential utilized today.
As the idea is to specify a size limit set by the condition that material
not be spilled from a star, the descriptive term 'limiting
lobe' serves well. The insight that spawned the concept came from \citet{kuiper},
who realized that tidal force is an unnecessary complication with regard to the lobe size limit,
as only one point, not two, need be considered, and only ordinary effective gravity at that
point, not differences between two points, need be computed. The procedure is (step 1) to find the point along
the line of centers where ordinary gravitational (not differential tide raising) forces due to $M$
and $m$, along with local rotational force, add to zero. Material that is stationary in a frame that co-rotates 
with the star is not bound to the star at this special 
point, so an ejection nozzle forms. Asynchronous examples are now treated via a factor $F^2$ (see \S\ref{loberad}) that 
alters the centrifugal term without affecting the basic idea of locating the effective gravity null point \citep{plavec58,limber}. A definite
equipotential that defines the star surface passes through the special point of null effective gravity, so (step 2) numerically integrate
the volume, $V_{lobe}$, enclosed by that equipotential and thereby find the equivalent-sphere mean radius as
\begin{equation}
R_{mean}=(3V_{lobe}/{4\pi})^{1/3}. \label{rlobe}
\end{equation}
Kuiper assumed synchronous rotation, which is the expected
and observationally indicated case for very close binaries (due to tidal locking).

Conditions that lead to small limiting lobes for SMBH orbiters are the enormous mass ratio, large 
orbital eccentricity, and -- not previously emphasized in the literature -- fast rotation. 
For fixed SMBH mass\footnote{We adopt $M_{SMBH}= 4.31\times 10^6$ $M_{\odot }$ \citep{gill09} and thus a mass
ratio, $M_{SMBH}/M_{10}$, of $4.31\times 10^5$ for a $10$ $M_{\odot }$ star.}, lobe size
decreases with decreasing star mass, decreasing orbit size, increasing
eccentricity, and increasing star rotation. 
S2 is known to be a fast rotator, while any kind of tidal locking would produce exceedingly
slow rotation in view of the 15.8 year \citep{ghezduch} orbit period, $P_{orb}$. 
The fastest locked rotations would be for locking to the pericenter orbital angular rate and give $P_{rot}$ 
around half a year for S2, whereas the \textit{measured} $V_{rot} \sin i$ is $220\pm40$ $km$ $sec^{-1}$ \citep{ghezduch},
so there is no tidal locking of any kind.
Whether the star's equator is aligned with the orbit plane in not known but the orbital
$\sin i$ is $0.7040\pm0.0058$ \citep{gill09} so, under the assumption of alignment, $V_{rot} \approx312\pm57$ $km$ $sec^{-1}$. 
The corresponding angular rotation, assuming $R_{eq}=5.0 R_\odot$, is $\approx 7100$ times the mean orbital angular rate, and 
rotational force becomes important in setting local effective gravity.
It will be shown below that the problem of main sequence SMBH orbiters being too small to exceed their limiting 
lobes can disappear if they rotate at typical B star rates, as does S2.

\subsection{The Eggleton Approximation}

Approximation formulas are often used to estimate lobe size, most commonly one by \citet{egg83}
that reproduces accurately computed mean lobe radii, based on the Kuiper logic, to better than 1 percent
over the full range of mass ratio, from $0$ to $\infty$\footnote{Incidentally, we checked the 1 percent accuracy statement 
in \citet{egg83} at the request of Prof. Eggleton, finding no discrepancies as large as $0.8$ 
percent among 18 widely spread mass ratios, of which only two exceeded half a percent. Column 2
of Eggleton's Table 1 (mean lobe radii from integrated volumes) was reproduced to all printed digits except
for two differences of 1 in the last place. Column 6, the Eggleton approximation, was
reproduced exactly. The checks
were done with the WD \citep{wd71} computer model.}. 
Although a rather small computer program can generate such lobe radii with
negligible error, and some public binary star programs list lobe radii
as incidental output, the Eggleton approximation has provided a one-line lobe calculation
in many evolutionary programs where 1 percent accuracy may be sufficient.
However note that the Eggleton formula is specifically for synchronous rotation and not meant
for stars that rotate faster or slower than synchronously. It will give limiting lobe radii
that are too large by orders of magnitude if applied to fast rotators such
as star S2, and may be responsible for misleading conclusions where rotation rates are unknown.
An algorithm that follows the Kuiper logic, enhanced to handle arbitrary rotation
and eccentricity, is not difficult to program and avoids the approximations of fitted formulas.
Most \textit{accurate} limiting lobe computations now adopt Kuiper's strategy, usually
via one of the commonly used binary system light/velocity curve programs, although
another option can be the collection of intricate approximation formulas by \citet{sepinski} that account for
asynchronism and eccentricity.

\section{Quantitative Estimate of S2's Lobe Size} \label{loberad}

Computation of a binary component's limiting lobe geometry begins with solution for a point of null effective
gravity along the line of star centers (x-axis), thereby locating the nozzle from which matter flows if the
lobe is filled or slightly overfilled. The relevant equation for the S2 problem must account for orbital
eccentricity and the star's rotation in addition to the gravity of both objects, 
as does eqn. 3 of \citet{wils79} for the derivative of potential\footnote{The potential is a modified
version according to the convention in \citet{kopal}.}
in the x-direction, which is zero at the null point. That equation is

\begin{equation}
\frac{d\Omega}{dx}=-\frac{x}{(x^2+y^2+z^2)^{3/2}} + \frac{q(D-x)}{([D-x]^2+y^2+z^2)^{3/2}} + F^2(1+q)x -q/D^2. \label{dodx}
\end{equation}

\noindent Rotation enters via a parameter $F$,
the ratio of rotational angular velocity $\omega_{rot}$ to mean (i.e. time-averaged) orbital angular velocity $\omega_{orb}$. 
Other input quantities are the component mass ratio ($q=M2/M1$), 
momentary separation of star centers ($D$), and $x,y,z$ rectangular coordinates of a point
at which $d\Omega/dx$, and subsequently $\Omega$, are to be computed. Here S2 is taken to be 
star 1 and the SMBH is object 2, so the mass ratio is a large number rather than its reciprocal.
The unit for $x$, $y$, and $z$ is $D$ while $D$ is in unit $a$, the semi-major axis of the 
relative orbit, in computations with equation \ref{dodx}, with $D=1-e$ at periastron or pericenter.
The location of the null point along the line of centers is found by setting $y=z=0$ and 
$d\Omega/dx$ also to $0$, setting the dimensionless angular rotation $F$ 
and eccentricity $e$ to values of interest, and then solving
for $x$ by numerical inversion (such as Newton-Raphson iteration).
The potential at the null point then establishes the 
lobe surface's 3-dimensional form as an equipotential that includes the null point 
(see eqn. 1 of \citet{wils79} for the generalized defining potential).
The equipotential's enclosed volume ($V$) can then be integrated numerically 
via the defining equation and a mean
lobe radius found from eqn. \ref{rlobe}.
A final step computes equatorial rotation velocity, $V_{eq}$, from angular velocity. That calculation is
simplified by the star being almost axially symmetric and its equator circular at these
fast rotation rates, so there is no issue of where along the equator the result applies.
Accordingly 

\begin{equation}
V_{eq}=R_{eq}\omega_{orb}F,
\end{equation}

\noindent with length in km, time in seconds, and
mean orbital angular velocity, $\omega_{orb}=2\pi/P_{orb}$, in $radians/sec$.

The binary star modeling and analysis program (WD program\footnote{The WD program's most recent public version,
with documentation and sample input files, can be downloaded from anonymous FTP site ftp.astro.ufl.edu.
Go to sub-directory pub/wilson/lcdc2013.}) applied here has refinements that allow reliable operation in difficult 
circumstances. For example, its Newton-Raphson iterations (for inversion of equation \ref{dodx}
to find the effective gravity null point) evaluate several Taylor series terms beyond 
the usual first derivative term. This point is mentioned so that readers who may write their 
own inversion program to check our results are not disappointed by failed computations.
A relatively simple inversion scheme can converge well for ordinary mass ratios
but not for ultra-large mass ratios such as the $4.31\times 10^5$ of the present problem
Also important for fractionally tiny lobes (large $q$, large $F$) 
is to begin iterations already close to the null point, so as to avoid an initial jump \textit{beyond}
the proper range between the star centers, from which recovery is difficult. Fortunately such a
configuration admits particularly good starting estimates of the null point's location. To see this readily,
write eqn. \ref{dodx} as it applies along the line of centers at the null point,

\begin{equation}
0= -\frac{1}{x^2} + \frac{q}{(D-x)^2} +F^2 (1+q) x -\frac{q}{D^2}.
\end{equation}

\noindent This form is a quintic equation in $x$, soluble only iteratively, but with $x$ very small the 
second and fourth terms on the right side very nearly cancel so that the remaining terms (also replacing 1+$q$ with the very large $q$) lead to
a simple result,

\begin{equation}
x\approx F^{-2/3} q^{-1/3}.
\end{equation}

\noindent The approximation is reasonably accurate only for quite small $x$, although very accurate for SMBH orbiters
and perhaps usefully accurate for $M_2/M_1$ of a few hundred or more. 

Inputs to the lobe size computation for S2 were $e=0.88$ \citep{gill09} and $M_{SMBH}/M_{S2}=4.31\times 10^5$ 
(mass ratio), along with a few well spaced $F$'s. One of the $F$'s
is close to the nominal value of 7100 that goes with our rough estimate of $V_{eq}$ that assumed alignment 
of the equatorial and orbit planes in \S\ref{lobesize}.
The resulting mean lobe radius is $6.5 R_\odot$, which is larger
than a 10 $M_\odot$ main sequence star (about 3 $R_\odot$ on the ZAMS to 5 $R_\odot$ at the TAMS), although  
the spectral type estimate by \citet{eisenhauer} extends to O8, for which a main sequence radius can exceed $6.5 R_\odot$.
A stripped highly evolved star that \textit{resembles} a main sequence star, as in \citet{davies}, remains a candidate.
With either kind of star, the idea of lobe overflow at pericenter passage now becomes a real possibility. Table \ref{lobe} 
has mean lobe radii\footnote{Note that these are 'equivalent sphere' radii, not 
distances to the effective gravity null point.} for four assumed angular rotation velocities (F's) of the 10 $M_\odot$ model orbiter to give a sense
of how steeply lobe size depends on rotation rate. A check to see if the program gives the 
right order of lobe size is provided by calculation of the equatorial radius of a 10 $M_\odot$ isolated star
(no SMBH) that is marginally unbound at the equator while rotating at one of the table 
values, $307$ $km$ $sec^{-1}$. If the magnitudes of rotational and gravitational force are then equated,
the equatorial radius will be given by $R_{eq}=GM/V^2$, which evaluates to $20.2 R_{\odot}$ for a 10 $M_\odot$ star. The corresponding \textit{mean} 
radius will be smaller since $R_{pole}$ is smaller than $R_{eq}$, so rotation alone produces a 
limiting size only about three times greater than do the combined effects of rotation and the SMBH gravity.
The purely gravitational lobe radius for a slowly rotating star, with $e=0.88$ and the present problem's adopted masses, is $\approx 100 R_{\odot}$, so the effect of fast 
rotation on lobe size is not small.

\subsection{Why Such Large Scale Ejection?}

A remaining issue is why a \textit{huge} puff would be ejected at pericenter passage. 
The ordinary context of lobe overflow is the synchronous-circular case that is commonly encountered
in close binary systems, where gas leaks out quiescently and is usually difficult or impossible
to detect photometrically. S2, being a very fast rotator, will not undergo the gentle process of the synchronous-circular case with its
low ejection velocity. The supersynchronous case is very different, with an ejection velocity close to the 
star's equatorial velocity, which is of order 300 $km$ $sec^{-1}$ for our model of S2.
And why would a large amount of gas be ejected? Suppose the \cite{davies} 
proposal, that the close-in orbiters are tidally stripped highly evolved stars, is correct, and that S2 is typical. Well known (e.g. \citet{plavec68})  
is that radii of highly evolved (i.e. chemically stratified) stars increase with loss of envelope matter,
in contrast with shrinkage for unevolved and modestly evolved stars. S2 has 15.8 years
between pericenter passages to expand following each pass and could arrive at pericenter 
not just marginally filling its lobe but substantially overfilling it. 
Although a quantitative estimate
of the overfilling will require reasonably good estimates of S2's internal structure that are
not now in hand, the qualitative picture is that S2 may reach pericenter ready to send very fast 
moving gas through a large open nozzle, leading to a very large ejection event. One test of this idea,
waiting for the next pericenter passage, is that emission lines should appear as the ejected gas expands and becomes
optically thin. Naturally some or all of these expectations may be anticipated as other SMBH orbiters 
pass through their pericenters.

\begin{deluxetable}{ccc}
\tablecaption{Dependence of 'Equivalent Spherical Volume' Lobe Radius on Equatorial Velocity for a 10 $M_\odot$, $e=0.88$ SMBH Orbiter Representing S2 \label{lobe}}
\tablehead{\colhead{F (rotation parameter)} & \colhead{Equatorial Velocity ($km\cdot sec^{-1}$)} & \colhead{'Equivalent Sphere' Lobe Size ($R_\odot)$}}
\startdata
1000  & 44 & 23.6 \\ 
3000  & 131  &  11.4  \\
7000  &  307  &  6.5  \\
10000  & 438  &  5.1  \\
\enddata
\end{deluxetable}

\end{document}